\documentstyle[12pt]{article}
\begin{document}
\begin{titlepage}
\title{Comment on 'Reduced System Dynamics from the N-Body Schr\"odinger
        Equation'}
\author{Lajos Di\'osi\\
        KFKI Research Institute for Particle and Nuclear Physics\\
        H-1525 Budapest 114, POB 49, Hungary\\
        e-mail: diosi@rmki.kfki.hu\\\\
        {\it e-Archive ref.: quant-ph/9509020}\\\\}
\date{September 29, 1995}
\maketitle
\begin{abstract}
We argue that the "reduced wave function", proposed recently
[Phys.Rev.Lett. {\bf 75}, 2255 (1995)], contains conditional
and restricted information on the reduced system. The concept
of "reduced wave function" can thus not represent a relevant
alternative to the common reduced dynamics methods.

\end{abstract}
\end{titlepage}

In his Letter [1], Muriel presents a time-dependent wave function for
one particle "in the company of $N-1$ other particles". Starting from
the N-particle wave function $\psi_N(r_1,r_2,\dots,r_N)$, the one-particle
wave function is derived by the projection $\psi_1=P\psi_N$ which,
in detailed form, reads:
$$
\psi_1(r_1;t)=
(1/\Omega^{N-1})\int\psi_N(r_1,r_2,\dots,r_N;t)dr_2 dr_3\dots dr_N.
$$
Assuming the standard N-body Schr\"odinger equation for $\psi_N$, the
time-dependent solution of the "reduced" one-particle wave function can
formally be written as
$$
\psi_1(t)=P\exp(-itH)\psi_N(0).
$$
with the N-body Hamiltonian $H$. The Letter calls the dynamics of $\psi_1$
"reduced dynamics" of the distinguished particle.

Here I would not discuss the Letter's main goal that is giving the
above solution $\psi_1(t)$ a more instructive form. I should, however,
question whether the Letter's "reduced dynamics" has enough to do with
the usual concept of reduced dynamics. Obviously, the projection $P$
projects the N-body quantum state onto the subspace where each body
except for the distinguished one is in {\it zero-momentum eigenstate}.
Consequently, the "reduced dynamics" of the Letter offers {\it conditional}
predictions for the distinguished particle while, unfortunately,
the conditions concern the {\it other} $N-1$ bodies. Calculating the
"reduced one-body wave function" $\psi_1(r_1,t)$ for $t>0$, one can predict
the expectation value of a Hermitian observable $A_1(r_1,r_1^\prime)$
of the distinguished body in the form
$$
{\int\overline{\psi}_1(r_1;t)A_1(r_1,r_1^\prime)
                           \psi_1(r_1^\prime;t)dr_1dr_1^\prime\over
 \int\overline{\psi}_1(r_1;t)\psi_1(r_1;t)dr_1}
$$
{\it provided the $N-1$ accompanying bodies are simultaneously found in
zero-momentum eigenstates}. In such a way, the set of theoretical
predictions which Muriel's "reduced dynamics" is capable to offer
becomes extremely restrictive.

One-body reduced dynamics, as it is commonly understood, are capable to
predict the expectation values of one-body observables $A_1$ {\it without
further tests on the accompanying bodies}. Consequently, Muriel's proposal
does not serve as true reduced dynamics and numerous theoretical
implications, claimed in the Letter, are not likely to be relevant for
the standard issue.

This work was supported by OTKA Grants No.1822/1991 and T016047.

\end{document}